\documentclass[vci-paper]{vciart}
\usepackage{amssymb}

\usepackage{graphicx}

\begin{document}

\begin{frontmatter}

\title{The K2K SciBar Detector}



\author[KEK]{K.~Nitta},
\author[BAR]{E.~Aliu},
\author[BAR]{S.~Andringa},
\author[KOB]{S.~Aoki},
\author[SNU]{S.~Choi},
\author[ROM]{U.~Dore},
\author[BAR]{X.~Espinal},
\author[VAL]{J.~J.~Gomez-Cadenas},
\author[WU]{R.~Gran},
\author[KYO]{M.~Hasegawa},
\author[KYO]{K.~Hayashi},
\author[HIR]{K.~Hayashi},
\author[KEK]{Y.~Hayato},
\author[KYO]{K.~Hiraide},
\author[KEK]{A.~K.~Ichikawa},
\author[HIR]{M.~Iinuma},
\author[CNU]{J.~S.~Jang},
\author[SNU]{E.~J.~Jeon},
\author[SNU]{K.~K.~Joo},
\author[SUNY]{C.~K.~Jung},
\author[KYO]{I.~Kato},
\author[SUNY]{D.~Kerr},
\author[CNU]{J.~Y.~Kim},
\author[SNU]{S.~B.~Kim},
\author[SUNY]{K.~Kobayashi},
\author[HIR]{A.~Kohara},
\author[KYO]{J.~Kubota},
\author[INR]{Yu.~Kudenko},
\author[OSK]{Y.~Kuno},
\author[CNU]{M.~J.~Lee},
\author[SUNY]{E.~Lessac-Chenin},
\author[CNU]{I.~T.~Lim},
\author[ROM]{P.~F.~Loverre},
\author[ROM]{L.~Ludovici},
\author[KYO]{H.~Maesaka},
\author[ROM]{C.~Mariani},
\author[SUNY]{C.~McGrew},
\author[INR]{O.~Mineev},
\author[KYO]{T.~Morita},
\author[KEK]{T.~Murakami},
\author[HIR]{Y.~Nakanishi},
\author[KYO]{T.~Nakaya},
\author[HIR]{S.~Nawang},
\author[KYO]{K.~Nishikawa},
\author[DSU]{M.~Y.~Pac},
\author[SNU]{E.~J.~Rhee},
\author[BAR]{A.~Rodr\'\i guez},
\author[BAR]{F.~Sanchez},
\author[KYO]{T.~Sasaki},
\author[WU]{K.~K.~Shiraishi},
\author[KOB]{A.~Suzuki},
\author[HIR]{T.~Takahashi},
\author[OSK]{Y.~Takubo},
\author[KEK]{M.~Tanaka},
\author[SUNY]{R.~Terri},
\author[VAL]{A.~Tornero-Lopez},
\author[KYO]{S.~Ueda},
\author[WU]{R.~J.~Wilkes},
\author[KYO]{S.~Yamamoto},
\author[KYO]{M.~Yokoyama}, and
\author[OSK]{M.~Yoshida}

\address[KEK]{Institute of Particle and Nuclear Studies, KEK, Tsukuba, Ibaraki 305-0801, JAPAN}
\address[BAR]{   Universitat Aut\'{o}noma de Barcelona Institut de F\'\i sica d'Altes Energies Edifici Cn, E-08193, Barcelona, SPAIN}
\address[KOB]{Kobe University, Kobe, Hyogo 657-8501, JAPAN}
\address[SNU]{Department of Physics, Seoul National University, Seoul 151-742, KOREA}
\address[ROM]{Universita' di Roma "La Sapienza" and INFN Roma, ITALY}
\address[VAL]{IFIC -- Instituto de Fisica Corpuscular Apartado de Correos 22085 E-46071 Valencia, SPAIN}
\address[WU]{Department of Physics, University of Washington, Seattle, WA 98195-1560, USA}
\address[KYO]{Department of Physics, Kyoto University, Kyoto 606-8502, JAPAN}
\address[HIR]{Graduate School of Advanced Sciences of Matter, Hiroshima University, Higashi-Hiroshima, 739-8530, JAPAN}
\address[CNU]{Department of Physics, Chonnam National University, Kwangju 500-757, KOREA}
\address[SUNY]{Department of Physics and Astronomy, State University of New York, Stony Brook, NY 11794-3800, USA}
\address[INR]{Institute for Nuclear Research RAS, 117312 Moscow, RUSSIA}
\address[OSK]{Department of Physics, Osaka University, Toyonaka, Osaka 560-0043, JAPAN}
\address[DSU]{Department of Physics, Dongshin University, Naju 520-714, KOREA}
                                                                                                                                                
\begin{abstract}
A new near detector, SciBar, for the K2K long-baseline neutrino oscillation experiment was installed
to improve the measurement of neutrino energy spectrum and to study neutrino interactions in the energy region around 1 GeV.
SciBar is a 'fully active' tracking detector with fine segmentation 
consisting of plastic scintillator bars.
The detector was constructed in summer 2003 and is taking data since October 2003.
The basic design and initial performance is presented.

\end{abstract}

\end{frontmatter}

\section{Introduction}
The KEK-to-Kamioka long-baseline neutrino experiment (K2K)\cite{k2k-detector} 
started taking data in 1999. 
An almost pure muon neutrino beam with average energy of 1.3~GeV 
is produced with the KEK 12-GeV proton synchrotron 
and directed toward the Super-Kamiokande detector (SK) located at Kamioka, 
250~km away from KEK.
The neutrino flux and energy spectrum at SK is estimated 
from the flux measured by near detectors located 300~m downstream 
from the production target. 
The number of events and the spectral shape at SK are compared with the expectations
to study neutrino oscillations.
The latest K2K results\cite{skatm} indicate neutrino oscillation, 
and are consistent with the SK results\cite{k2k-osci}.
In K2K,
the neutrino energy at the oscillation maximum is expected to be $\sim$0.6~GeV.

A new near detector, SciBar ($\underline{\textrm{Sci}}$ntillator $\underline{\textrm{Bar}}$), 
was constructed in Summer 2003 to upgrade the near detectors.
A main motivation of the new detector is to improve the measurement of
neutrino energy spectrum by using Charged Current Quasi Elastic interaction 
(CCQE, $\nu_\mu+n\rightarrow\mu^-+p$).
In order to select CCQE interaction with high purity and high efficiency and
to suppress other interactions such as 
inelastic interactions with pions ($\nu_\mu+p\rightarrow\mu^-+p+\pi^+$), 
the detector is designed to have high efficiency for all charged particles
produced in the interaction.
The detector consists of plastic scintillator strips with fine segmentation.
The scintillator itself is a neutrino target and has no dead region.
Due to the fine segmentation of the detector, 
short tracks down to 10~cm long can be detected. 
The detector also has a capability of particle identification
(especially for protons and pions) with dE/dx information by measuring the energy deposit in each strip.

In addition, we expect to provide many new results for neutrino interactions
in the 1~GeV region with SciBar.

\section{The SciBar Detector}
\begin{figure}
\centering
\includegraphics[scale=0.8]{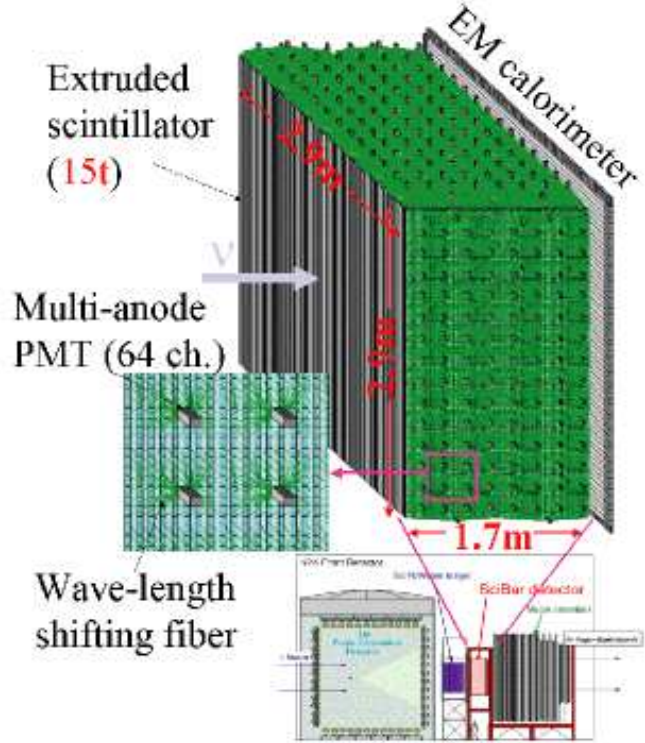}
\caption{Schematic drawings and description of the SciBar detector.}
\label{fig:scibar}
\end{figure}

Figure~\ref{fig:scibar} shows a schematic view of SciBar.
SciBar consists of two components:
a tracking calorimeter made of scintillator strips 
and an electromagnetic calorimeter called Electron Catcher.
Table \ref{table:structure} shows the design for each component.
\begin{table}
\centering
\scalebox{0.8}{
\begin{tabular}[t]{lll}
\hline\hline
Structure& Dimensions & 2.9~m (horizontal), 2.9~m (vertical), 1.7~m (thick)\\
	& Weight    & 15~tons \\
	& Number of Strips & 14,848 \\
	& Number of PMTs &  224 (+OD:8)\\
\hline
Scintillator   & Material & polystyrene, PPO(1\%) and POPOP(0.03\%)\\
strip	& Size & 1.3~cm (thick), 2.5~cm (wide), 3~m (long)\\
         & Weight     & 1~kg\\
	 & Coating    & 0.25~mm (TiO$_2$) \\
	 & Emission length & 420 nm (peak) \\
\hline
Fiber		& Diameter	& 1.5~mm$\phi$ \\
	& Reflective index & 1.59 (outer) / 1.50 (middle) / 1.42 (inner)\\
	& Absorption length & 430 nm (peak) \\
	& Emission length & 476 nm (peak) \\
\hline
PMT     & Model & H8804 \\
	& Cathode material & Bialkali \\
        & Anode  & 8 $\times$ 8 (2$\times$2mm$^2$/pixel)\\
        & Wave length sensitivity  & 300-650nm (Max 420nm) \\
	& Number of dynode stage & 12 \\
	& Gain(@800V) & 3$\times10^5$ \\
	& Quantum efficiency & 21 \% at 390nm \\ 
\hline
DAQ   	& Shaping time & 80 nsec (TA), 1.2$\mu$sec (VA)\\
	& Pedestal width & 0.03 MeV \\
        & Linearity & 5 \% at 30 MeV\\
	& Dynamic range & 0.1-30~MeV\\
\hline\hline
\end{tabular}
}
\caption{Specification of each component of SciBar.}
\label{table:structure}
\end{table}
The SciBar tracker consists of 14848 extruded scintillator strips
with each dimension of 1.3$\times$2.5$\times$300~cm$^3$.
The scintillator strips are arranged in 64 layers.
Each layer consists of two planes, with 116 strips glued together
to give horizontal and vertical position.
The total size and weight are 2.9$\times$2.9$\times$1.7~m$^3$ and 15~tons, respectively.
Each strip is read out by a wavelength shifting (WLS) fiber attached to a 64-channel multi-anode PMT (MAPMT). 
Charge and timing information from MAPMT is recorded by custom designed electronics\cite{daq}.
Two outermost strips in each horizontal and vertical planes 
are called as Outer Detector (OD).
For the OD readout, eight 1-ch PMTs with WLS fibers are used 
to identify incoming and outgoing events.

The scintillator strips are made of polystyrene, infused with PPO (1~\%) and POPOP (0.03~\%), 
and are produced by extrusion in the shape of rectangular bar 
with TiO$_2$ reflecting coating (0.25~mm thickness).
Scintillators are produced at FNAL
and the 
composition of the material is the
same as those used 
for the MINOS experiment\cite{minos}. 
Each scintillator has a 1.8~mm diameter hole
where the 1.5~mm$\phi$ WLS fiber is inserted for light collection.
The fiber is multi-clad type, Y11(200)MS, made by Kuraray.
The attenuation lengths of all WLS fibers were measured 
before the installation.
Their average is 350~cm.

MAPMT is H8804 made by Hamamatsu Photonics K.K..
It's Anodes are arranged in an 8 $\times$ 8 array 
with each anode measuring 2 mm $\times$ 2 mm.
The sensitive wave length is from 300~nm to 650~nm, 
which matches the emission spectrum of the WLS fibers.
The cross talk value in our configuration is less than 4 \%.
The gains of all channels have been measured before installation.
The linearity is also measured to be 10~\% upto 300~photo-electrons~(p.e.)
with the gain of $6.3\times10^5$.

Readout system consists of frontend electronics 
attached to MAPMT and a backend VME module.
A combination of ASICs (VA32HDR11 and TA32CG made by IDEAS) is used on the frontend readout electronics.
VA is a 32-ch preamplifier chip with shaper and multiplexer.
TA provides timing information after taking OR of 32 channels.
Two packages of VA/TA are mounted on a custom-designed PCB board (Front-End board, FEB) to read out signals from 64 anodes.
A backend electronics board (DAQB) has been also newly developed as a standard VME-9U board.
DAQB controls and reads out eight FEBs.
The charge information from MAPMT is digitized with a 12-bit flash ADC and read out through VME bus.
The timing information is processed and recorded by a TDC.
We use a 64-ch, multi-hit TDC developed for ATLAS muon chamber.
In order to separate of protons from pions by the energy deposit,
the wide dynamic range is required for the readout system.
The system is designed to have a good linearity up to 300~p.e., 
corresponding to 30~MeV, while
the pedestal width is 2 ADC counts,
corresponding to 0.03 MeV.

Electron catcher (EC) 
is located downstream of the tracker.
It is used to measure the $\nu_e$ contamination in the beam
and $\pi^0$ production events.
The EC is a calorimeter
consists of two planes of 30 (horizontal) and 32 (vertical) modules 
(4 $\times$ 8 $\times$ 262~cm$^3$) re-used from the CHORUS experiment\cite{chorus}.
Each plane has thickness 5.5 radiation length. 
A module of 4 $\times$ 4~cm$^2$ cell consists of 1 mm diameter scintillating fibers 
in the grooves of lead foils.
Lights from both ends of the module are read out by two PMTs.
Energy resolution is $14/\sqrt E$(GeV)~\%
with linearity better than 10~\%.
The EC is used for electron and photon identification and the measurement of their energy.

\section{Basic performance}
\begin{figure}
 \begin{minipage}[t]{.47\textwidth}
 \includegraphics[scale=0.29]{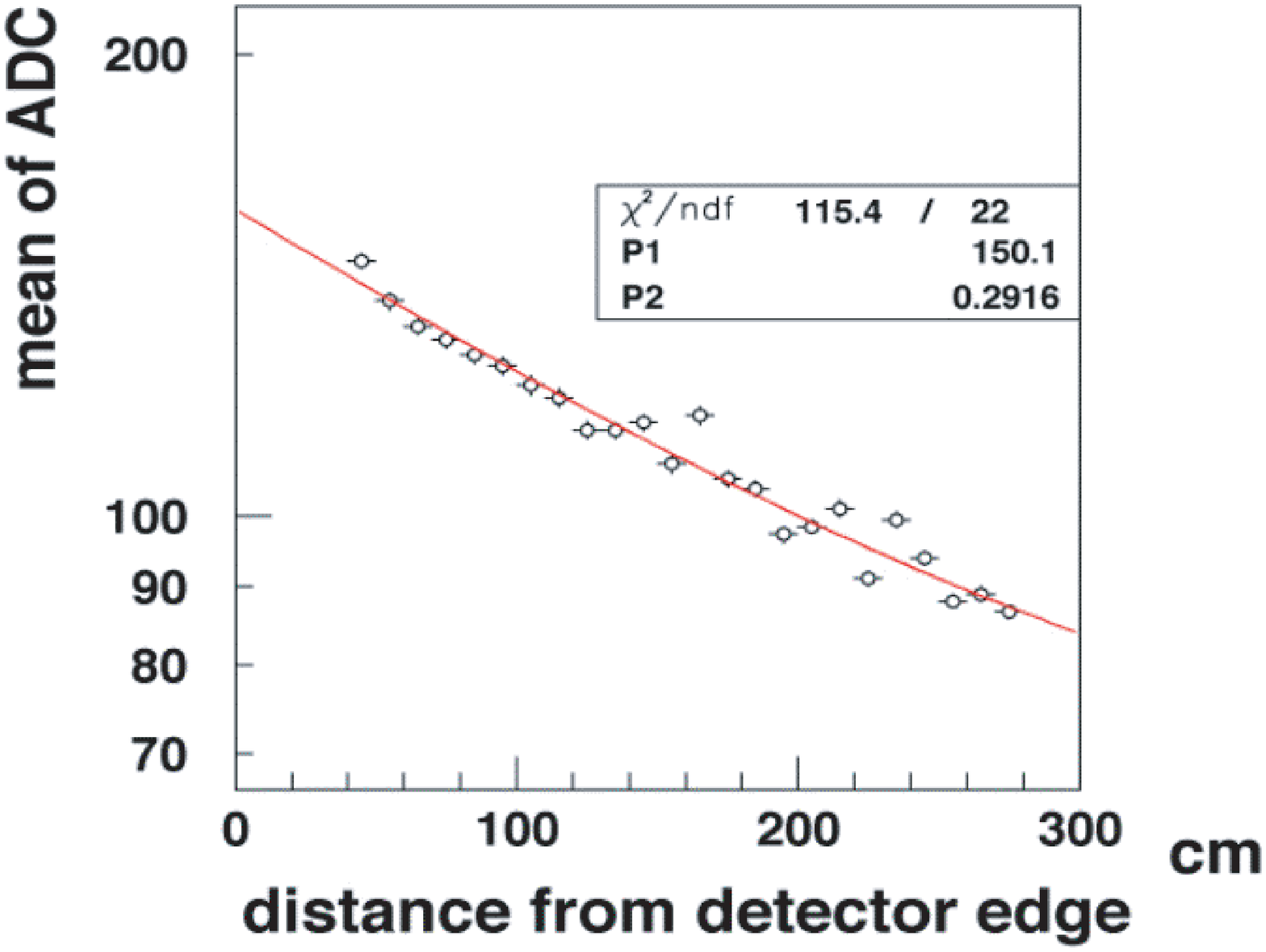}
 \caption{Attenuation curve of a WLS fiber obtained by the measurement of cosmic-rays.}
 \label{fig:wlsatt}
 \end{minipage}
\hfill
 \begin{minipage}[t]{.47\textwidth}
 \centering
 \includegraphics[scale=0.6]{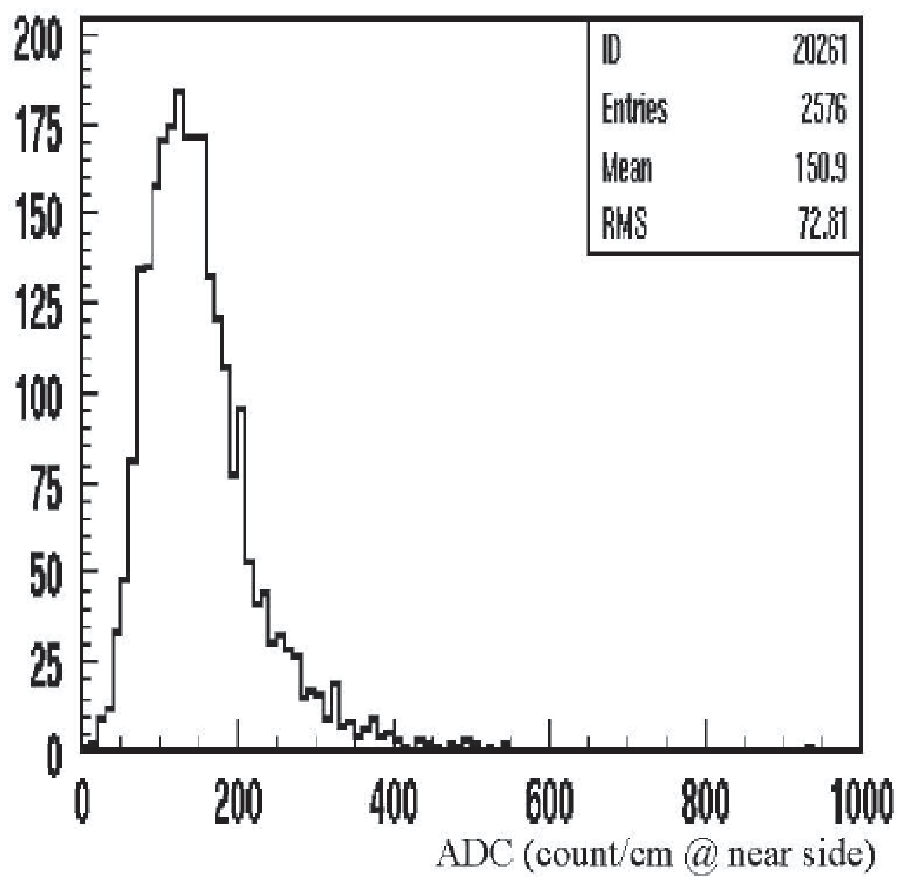}
 \caption{Pulse height distribution for a strip by the measurement with cosmic ray.
The effect of attenuation in a WLS fiber is corrected.}
 \label{fig:energycalib}
 \end{minipage}
\end{figure}

\begin{figure}
\centering
\includegraphics[scale=0.7]{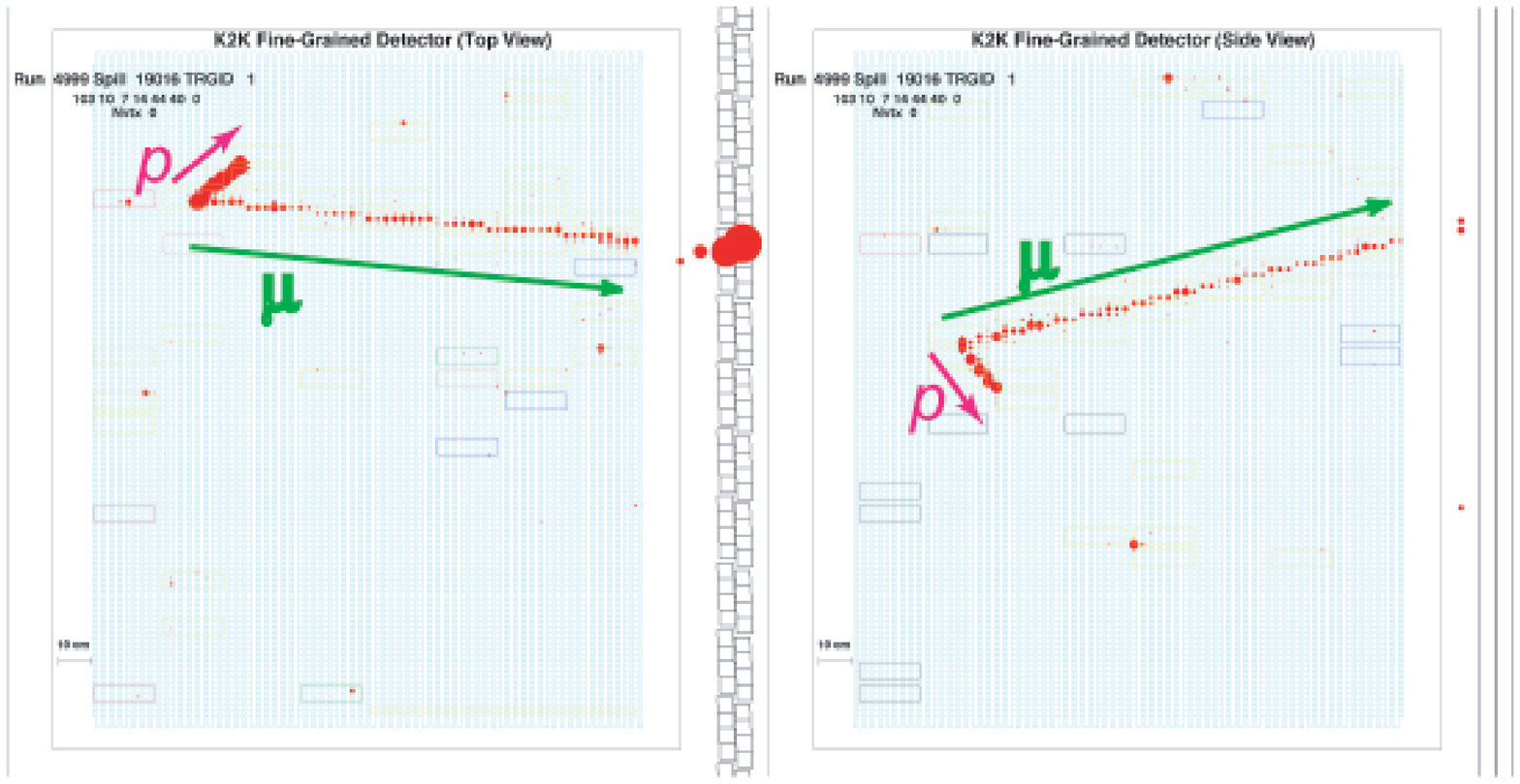}
\caption{Event displays of a CCQE candidate
in top view (left) and side view (right).
Each red circle corresponds to a hit in SciBar. 
Area of the circle is proportional to the ADC count (i.e. energy deposit) of 
each channel.
Boxes represent TDC hit information by their colors.
}
\label{fig:disp-CCQE}
\end{figure}

SciBar is taking data since October 2003.
Pedestal, LED and cosmic-ray data are taken simultaneously with beam data
for calibration and monitoring.
The gain is monitored and corrected by LED, so
the energy deposit is measured within a stability better than 1 \% by the gain correction.
By using cosmic-ray data 
the attenuation length of WLS fibers are measured
as shown in Fig.~\ref{fig:wlsatt}.
The average attenuation length is 350~cm
which is consistent with the measurement before installation.
Figure \ref{fig:energycalib} shows the pulse height distribution 
by cosmic-ray muons in one strip. 
Averaging over all strips, the light yield is 16.5~p.e./cm for 
minimum ionizating particles at the detector edge close to PMT.
The timing resolution is 1.3 nsec with the TQ correction.
Neutrino data are taken every 2.2~seconds of beam cycle.
The expected number of neutrino interactions is 7$\times$10$^4$ events in a year.
The performance of SciBar is summarized in Table \ref{table:performance}.
Figure \ref{fig:disp-CCQE} shows the display of a CCQE candidate.
\begin{table}
\centering
\scalebox{0.8}{
\begin{tabular}[t]{ll}
\hline\hline
Trigger threshold & 0.1 MeV/strip\\
Attenuation length & 350cm \\
Timing resolution & 1.3 nsec\\
Energy resolution & a few \%(muon)*, $\sim$10 \%(electron)*\\
Number of neutrino interaction & 7.0$\times 10^4$ events/year (3$\times 10^{19}$ POT**)*\\
P/$\pi$ misidentification & 5\% ($<$1.0~GeV/c proton)* \\
\hline\hline
$*$ : Design value 
$ **$: Proton On Target
\end{tabular}
}
\caption{Performance of SciBar}
\label{table:performance}
\end{table}
We can clearly identify a proton track and a muon by their energy deposit.
In one year, 3$\times$10$^4$ CCQE events are expected to be observed.
SciBar works stably for four months of operation
and collects various events of neutrino interactions.

\section{Summary}
The SciBar detector is designed and constructed 
to measure the neutrino energy spectrum around 1 GeV
as the near detector of the K2K experiment.
,
and has been successfully commissioned
on schedule.
Now SciBar works stably with good performance.

\section{Acknowledgement}
We gratefully acknowledge the assistance of T.~Haff and P.~Rovegno 
during the construction of the detector. 
We would like to express our appreciation for support of M.~Taino during the construction work.
This work has been undertaken with the support of
the Ministry of Education, Culture, Sports, Science and Technology, Government of Japan 
and its grants for
Scientific Research, the Japan Society for Promotion of Science, the U.S. Department of Energy, the Korea Research Foundation, 
and the Korea Science and Engineering Foundation.
From the spanish institutes, we would like to thank the support of the spanish ministry of science.

\end{document}